\documentclass{article}
\pdfoutput=1

\usepackage{amsmath,amssymb,array,multirow,tikz,tikz-cd}

\usepackage{amsthm}
\theoremstyle{definition}
\newtheorem{definition}{Definition}

\theoremstyle{theorem}
\newtheorem{proposition}{Proposition}
\newtheorem{lemma}{Lemma}
\newtheorem{theorem}{Theorem}

\theoremstyle{definition}

\newtheorem{fact}{Fact}

\usetikzlibrary{arrows,automata}

\usepackage{hyperref}

\newcommand{\MBI}{\textsf{I}}
\newcommand{\MBE}{\textsf{E}}
\newcommand{\MBJ}{\textsf{J}}
\newcommand{\MBP}{\textsf{P}}
\newcommand{\MBS}{\textsf{S}}
\newcommand{\MBT}{\textsf{T}}
\newcommand{\MBF}{\textsf{F}}
\newcommand{\MBN}{\textsf{N}}

\newcommand{\MBTI}{\mathbb{MB}}
\newcommand{\MBTIts}{\textrm{MBTI}}
\newcommand{\MBTIps}{\mathcal{MBTI}}

\newcommand{\ISTJ}{\textsf{ISTJ}}
\newcommand{\ISFJ}{\textsf{ISFJ}}
\newcommand{\INFJ}{\textsf{INFJ}}
\newcommand{\INTJ}{\textsf{INTJ}}
\newcommand{\ISTP}{\textsf{ISTP}}
\newcommand{\ISFP}{\textsf{ISFP}}
\newcommand{\INFP}{\textsf{INFP}}
\newcommand{\INTP}{\textsf{INTP}}
\newcommand{\ESTP}{\textsf{ESTP}}
\newcommand{\ESFP}{\textsf{ESFP}}
\newcommand{\ENFP}{\textsf{ENFP}}
\newcommand{\ENTP}{\textsf{ENTP}}
\newcommand{\ESTJ}{\textsf{ESTJ}}
\newcommand{\ESFJ}{\textsf{ESFJ}}
\newcommand{\ENFJ}{\textsf{ENFJ}}
\newcommand{\ENTJ}{\textsf{ENTJ}}

\newcommand{\mbbb}{-!!!}
\newcommand{\mbb}{-!!}
\newcommand{\mb}{-!}
\newcommand{\m}{-}
\newcommand{\n}{0}
\newcommand{\p}{+}
\newcommand{\pb}{+!}
\newcommand{\pbb}{+!!}
\newcommand{\pbbb}{+!!!}
\newcommand{\pmlb}{\pm_{!}}
\newcommand{\pmub}{\pm^{!}}

\newcommand{\F}[2]{\mbox{$\textsf{#1}#2$}}

\newcommand{\signatures}{\mathbb{S}}
\newcommand{\factors}{\mathbb{F}}

\newcommand{\SPPts}{\textrm{SPP}}
\newcommand{\SPPps}{\mathcal{SPP}}

\newcommand{\atoms}{\mathbb{A}}
\newcommand{\LPL}{\textrm{LPL}}
\newcommand{\LPLbis}{\mathcal{LPL}}

\newcommand{\rightG}[1]{{#1^{\triangleright}}}
\newcommand{\leftG}[1]{{#1^{\triangleleft}}}
\newcommand{\rightleftG}[1]{{#1^{\triangleright\triangleleft}}}
\newcommand{\leftrightG}[1]{{#1^{\triangleleft\triangleright}}}
\newcommand{\rightI}{\mathrm{i}}
\newcommand{\leftI}{\mathrm{p}}

\newcolumntype{C}{>{\centering\arraybackslash}p{0.26\textwidth}}

\begin{document}
\title{A Galois-Connection between Myers-Briggs' Type Indicators and Szondi's Personality Profiles}
\author{Simon Kramer\\[\jot]
		\texttt{simon.kramer@a3.epfl.ch}}
\maketitle
\begin{abstract}
We propose a computable Galois-connection between 
	\emph{Myers-Briggs' Type Indicators (MBTIs),} 
		the most widely-used personality measure for non-psychi\-atric populations 
			(based on C.G.\ Jung's personality types), and 
	\emph{Szondi's personality profiles (SPPs),}  
		a less well-known but, as we show, finer personality measure for 
			psychiatric as well as non-psychiatric populations 
				(conceived as a unification of the depth psychology of S.\  Freud, C.G.\ Jung, and A.\ Adler).
The practical significance of our result is that 
	our Galois-connection provides a pair of computable, 
		interpreting translations between the two personality spaces of MBTIs and SPPs: 
			one \emph{concrete} from MBTI-space to SPP-space (because SPPs are finer) and
			one \emph{abstract} from SPP-space to MBTI-space (because MBTIs are coarser).
Thus Myers-Briggs' and Szondi's personality-test results are 
	mutually interpretable and inter-translatable, 
		even automatically by computers.
	
	\smallskip
	
\end{abstract}

\section{Introduction}
According to \cite[Page~xxi and 210]{GiftsDiffering},
	the \emph{Myers-Briggs Type Indicator (MBTI)} \cite{MyersBriggs},  
		based on C.G.\ Jung's personality types \cite{PsychologicalTypes}, has become 
			``the most widely-used personality measure for non-psychiatric populations'' and
			``the most extensively used personality instrument in history''
			 	with over two million tests taken per year.
In this paper,
	we propose a computable Galois-connection \cite{DaveyPriestley} between 
		MBTIs and 
		\emph{Szondi's personality profiles (SPPs)} \cite{Szondi:ETD:Band1}, 
			a less well-known but, as we show, finer personality measure for 
				psychiatric as well as non-psychiatric populations, and 
					conceived as a unification \cite{Szondi:IchAnalyse} of the depth psychology of 
						S.\ Freud, C.G.\ Jung, and A.\ Adler.

Our result 
	is a contribution to \emph{mathematical psychology} in the area of depth psychology, which 
		does not yet seem to have been explored with mathematical means  
			besides those of statistics (often not part of mathematics departments).
It is also meant as a contribution towards 
	practicing psychological research with the methods of 
		the exact sciences, for
			obvious ethical reasons.
The practical significance of our result is that 
		our Galois-connection provides a pair of efficiently computable, 
			interpreting translations between the two personality spaces of MBTIs and SPPs 
				(and thus hopefully also between their respective academic and non-academic communities): 
				one \emph{concrete} translation from MBTI-space to SPP-space (because SPPs are finer than MBTIs) and
				one \emph{abstract} translation from SPP-space to MBTI-space (because MBTIs are coarser than SPPs).
Thus Myers-Briggs' and Szondi's personality-test results are  
	mutually interpretable and inter-translatable, 
		even automatically by computers.
The only restriction to this mutuality is 
	the subjective interpretation of the faithfulness of these translations.
In our interpretation,
	we intentionally restrict the translation from SPP-space to MBTI-space, and only that one, 
		in order to preserve (our perception of) its faithfulness. 
More precisely,
	we choose to map some SPPs to the empty set in MBTI-space
		(but every MBTI to a non-empty set in SPP-space).
Our readers can 
	experiment with their own interpretations, 
		as we explain below.
		
We stress that 
	our Galois-connection between the spaces of MBTIs and SPPs is 
		independent of their respective \emph{test,} which 
			evaluate their testees in terms of 
				\emph{structured result values}---the MBTIs and SPPs---in the respective space.
Both tests are preference-based, more precisely, 
	test evaluation is based 
		on choices of preferred questions in the case of the MBTI-test \cite{MyersBriggs} and 
		on choices of preferred portraits in the case of the Szondi-test \cite{Szondi:ETD:Band1,SzondiTestWebApp}.
Due to the independence of our Galois-connection from these tests,
	their exact nature need not concern us here.
All what we need to be concerned about is the nature of the structured result values that these tests generate.
(Other test forms can generate the same form of result values, e.g.~\cite{Kenmo:Szondi}.)
We also stress 
	that our proposed Galois-connection is 
		what we believe to be an interesting candidate brain child for adoption by the community, but
	that there are other possible candidates, which our readers are empowered to explore themselves.
In fact,
	not only 
		do we propose a candidate Galois-connection between MBTI-space and SPP-space, but also 
		do we propose a whole \emph{methodology} for generating such candidates.
All what 
	readers interested in generating such connections themselves need to do is 
		map their own intuition about 
			the meaning of MBTIs to a standard interlingua, 
				called \emph{Logical Pivot Language (LPL)} here, and check that 
					their mapping has a single simple property,
						namely the one stated as Fact~\ref{fact:FactsAboutip}.1 about 
							our mapping $\rightI$ in  
								Figure~\ref{figure:MappingsAndMorphisms}.
Their desired Galois-connection is then automatically induced jointly by 
	their chosen mapping and 
	a mapping, called $\leftI$ here, from SPP-space to LPL that
		we choose once and for all possible Galois-connections of interest.
What is more,
	our methodology is applicable even more generally to the generation of Galois-connections between 
		pairs of result spaces of other personality tests.
SPPs just happen to have a finer structure than 
	other personality-test values that we are aware of, and 
		so are perhaps best suited to play 
			the distinguished role of explanatory semantics for result values of other personality tests.
Of course our readers are still free to choose their own preferred semantic space.
	
An SPP can be conceived as a tuple of eight, 
		so-called \emph{signed factors} whose signatures can in turn take twelve values.
So SPPs live in an eight-dimensional space.
On the other hand,
	an MBTI can be conceived as a quadruple of two-valued components, namely, 
		first, extro-/introversion, 
		second, perception, being either sensing or intuition, 
		third, judgment, being either thinking or feeling, and
		fourth, a dominance flag, indicating either a dominance of perception or judgment.
So MBTIs live in a coarser, four dimensional space.
Hence the translation from SPPs to MBTIs must be a projection (and thus surjection) of SPP-space onto MBTI-space.  
An insight gained in the finer referential system of SPPs is that  		
	MBTIs turn actually out to be non-orthogonal or not independent,
		contrary to common belief \cite{MyersBriggs,GiftsDiffering}.
Of course our readers are still free to disagree on the value of this insight by
	giving a convincing argument for why SPP-space would be an inappropriate semantics for MBTI-space.
After all,
	Szondi conceived his theory of human personality as 
		a unifying theory that also includes Jung's theory, 
			on which MBTI-theory is based.
We now put forward our own argument for why we believe SPP-space is indeed  
	an appropriate---though surely not the only---semantics for MBTI-space.
In Section~\ref{section:Structures},
	we present the defining mathematical structures for each space, and 
in Section~\ref{section:MappingsAndMorphisms},
	the defining mathematical mappings for their translation.
No prior knowledge of either MBTIs or SPPs is required to appreciate the results of this paper.

\section{The connection}
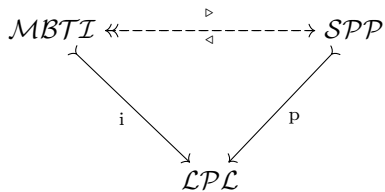
\begin{figure}
	\caption{Mappings between personality spaces and interlingua}
	$$\begin{tikzcd}
		\MBTIps \arrow[tail, swap]{ddr}{\rightI} \arrow[yshift=1ex, dashed]{rr}{\rightG{}} & & \SPPps \arrow[tail]{ddl}{\leftI} \arrow[dashed, yshift=-0.5ex, two heads, name=T, below]{ll}{\leftG{}}\\
		&&\\
		& \LPLbis &
	\end{tikzcd}$$
	\label{figure:MappingsAndMorphisms}
\end{figure}

In this section, 
	we present 
		the defining mathematical structures for 
			MBTI-space, the interlingua LPL, and SPP-space, as well as
		the defining mathematical mappings for 
			the concrete translation of MBTI-space to SPP-space and 
			the abstract translation of SPP-space back to MBTI-space, both via LPL, 
				see Figure~\ref{figure:MappingsAndMorphisms}.
				
\subsection{Structures}\label{section:Structures}
In this section, 
	we present 
		the defining mathematical structures for 
			MBTI-space, the interlingua LPL, and SPP-space.
We start with defining MBTI-space.
\begin{definition}[The Myers-Briggs Type Indicator Space]
Let 
	$$\MBTI=\{ \MBE, \MBI, \MBF, \MBT, \MBN,\MBS, \MBJ, \MBP \}$$
be the set of basic type indicators, with 
	\MBE\ meaning ``extroversion,'' 
	\MBI\ ``introversion,'' 
	\MBF\ ``feeling,'' 
	\MBT\ ``thinking,''
	\MBN\ ``intuition,''
	\MBS\ ``sensing,''  
	\MBJ\ ``judging,'' and
	\MBP\ ``perceiving.'' 
Further let 
	$$\begin{array}[t]{@{}r@{\ }c@{\ }l@{}}
		\MBTIts &=& \begin{array}[t]{@{}l@{\;}l@{}}
						\{&\ISTJ,\ISFJ,\INFJ,\INTJ,\ISTP,\ISFP,\INFP,\INTP,\\
						  &\ESTP,\ESFP,\ENFP,\ENTP,\ESTJ,\ESFJ,\ENFJ,\ENTJ\;\}
					\end{array}
		\end{array}$$
be the set of Myers-Briggs Type Indicators (MBTIs) \cite{MyersBriggs,GiftsDiffering}.

Then, 
	$$\MBTIps=\langle\, 2^{\MBTIts},\emptyset,\cap,\cup,\MBTIts,\overline{\,\cdot\,},\subseteq\,\rangle$$
defines our \emph{Myers-Briggs Type Indicator Space,} that is,
	the (inclusion-ordered, Boolean) powerset algebra \cite{DaveyPriestley} on \MBTIts\ 
		(the set of all subsets of \MBTIts).
\end{definition}
\noindent
Note that 
	we do need to define $\MBTIps$ as the set of all \emph{subsets} of $\MBTIts$ and 
		not simply as the set of all elements of $\MBTIts$.
The reason is the aforementioned fact that 
	in the finer referential system of SPP-space (see Definition~\ref{definition:SPP}), 
		MBTIs turn out to be non-orthogonal or not independent, and thus 
			an MBTI may have to be mapped to a proper set of SPPs (see Table~\ref{table:MBTItoLPL}).
So the proper setting for SPP-space is a set of \emph{subsets} of SPPs, which
	in turn, via the backward translation from SPP-space to $\MBTIps$, means that 
		the proper setting for $\MBTIps$, as the target of a mapping of subsets, 
			is also a set of subsets.
Further, 
	notice that 
		the MBTI-test \cite{MyersBriggs}, which as previously mentioned requires answering to questions, 
			actually requires 
				the $\MBP$-faculty (perception of the question), 
				the $\MBN$-faculty (intuition in the sense of textual, and thus symbolic understanding), and 
				the $\MBJ$-faculty (judgment in the sense of choice of and decision about an answer) 
					as its own prerequisites.
Incidentally, 
	the concept of choice is the key concept in Szondi's depth-psychological \emph{fate analysis}  \cite{Szondi:Schicksalsanalyse,Szondi:IchAnalyse}, which
	is the background theory for his test \cite{Szondi:ETD:Band1} and the SPPs that it generates.

We continue to define SPP-space.
\begin{definition}[The Szondi Personality Profile Space]\label{definition:SPP}
Let us consider the Hasse-diagram \cite{DaveyPriestley} in Figure~\ref{figure:SzondiSignatures} 
\begin{figure}[t]
\centering
\caption{Hasse-diagram of Szondi's signatures}
\medskip
\fbox{\begin{tikzpicture}
	\node (pbbb) at (0,4) {$+!!!$};
	\node (pbb) at (0,3) {$+!!$};
	\node (pb) at (0,2) {$+!$};
	\node (p) at (0,1) {$+$};
	\node (n) at (0,0) {$0$};
	\node (m) at (0,-1) {$-$};
	\node (mb) at (0,-2) {$-!$};
	\node (mbb) at (0,-3) {$-!!$};
	\node (mbbb) at (0,-4) {$-!!!$};
	\draw (mbbb) -- (mbb) -- (mb) -- (m) -- (n) -- (p) -- (pb) -- (pbb) -- (pbbb);	
 	\node (pmub) at (1,1) {$\pm^{!}$};  
	\node (pm) at (1,0) {$\pm$};
	\node (pmlb) at (1,-1) {$\pm_{!}$};
	\draw (pmlb) -- (pm) -- (pmub);
\end{tikzpicture}}
\label{figure:SzondiSignatures}
\end{figure}
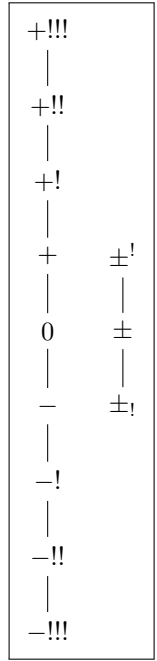
of the partially ordered set of \emph{Szondi's twelve signatures} \cite{Szondi:ETD:Band1} of 
human reactions, which are:
\begin{itemize}
	\item approval: from strong $+!!!$\,, $+!!$\,, and $+!$ to weak $+$\,; 
	\item indifference/neutrality: $0$\,; 
	\item rejection: from weak $-$\,, $-!$\,, and $-!!$ to strong $-!!!$\,; and 
	\item ambivalence: $\pm^{!}$ (approval bias), $\pm$ (no bias), and $\pm_{!}$ (rejection bias).
\end{itemize}
(Szondi calls the exclamation marks in his signatures \emph{quanta.})

Further let us call this set of signatures $\mathbb{S}$, that is,
		$$\signatures=\{\,\mbbb,\mbb,\mb,\m,\n,\p,\pb,\pbb,\pbbb,\pmlb,\pm,\pmub\,\}.$$	
	
Now let us consider \emph{Szondi's eight factors and four vectors} of 
	human personality \cite{Szondi:ETD:Band1} as summarised in Table~\ref{table:SzondiFactors}.
	\begin{table}[t]
		\centering
		\caption{Szondi's factors and vectors}
		\medskip
		{\small
		\begin{tabular}{|c|c||C|C|}
			\hline
			\multirow{2}{12.5ex}{\centering \textbf{Vector}} & \multirow{2}{7.5ex}{\textbf{Factor}} & \multicolumn{2}{c|}{\textbf{Signature}}\\
			\cline{3-4}
			&& $+$ & $-$\\
			\hline
			\hline
			\multirow{2}{12.5ex}{\centering \textsf{S} (Id)} & \textsf{h} (love) & physical love & platonic love\\
			\cline{2-4}
			& \textsf{s} (attitude) & (proactive) activity & (receptive) passivity\\
			\hline
			\multirow{2}{12.5ex}{\centering \textsf{P}\\[-\jot] (Super-Ego)} & \textsf{e} (ethics) & ethical behaviour & unethical behaviour\\
			\cline{2-4}
			& \textsf{hy} (morality) & immoral behaviour & moral behaviour\\
			\hline
			\multirow{2}{12.5ex}{\centering \textsf{Sch} (Ego)} & \textsf{k} (having) & having more & having less\\
			\cline{2-4}
			& \textsf{p} (being) & being more & being less\\
			\hline			
			\multirow{2}{12.5ex}{\centering \textsf{C} (Id)} & \textsf{d} (relations) & unfaithfulness & faithfulness\\
			\cline{2-4}
			& \textsf{m} (bindings) & dependence & independence\\
			\hline		
		\end{tabular}}
		\label{table:SzondiFactors}
	\end{table}
(Their names are of clinical origin and need not concern us here.)
And let us call the set of factors $\factors$, that is, 
	$$\factors=\{\,\F{h}{},\F{s}{},\F{e}{},\F{hy}{},\F{k}{},\F{p}{},\F{d}{},\F{m}{}\,\}.$$

Then,
	$$\SPPts=\{\; \begin{array}[t]{@{}l@{}}
				((\F{h}{,s_{1}}), (\F{s}{,s_{2}}), (\F{e}{,s_{3}}), (\F{hy}{,s_{4}}), 
				 (\F{k}{,s_{5}}), (\F{p}{,s_{6}}), (\F{d}{,s_{7}}), (\F{m}{,s_{8}})) \mid\\ 
				 s_{1},\ldots,s_{8}\in\signatures\;\}
				 \end{array}$$
is the set of Szondi's personality profiles, and
	$$\SPPps=\langle\, 2^{\SPPts},\emptyset,\cap,\cup,\SPPts,\overline{\,\cdot\,},\subseteq\,\rangle$$
defines our \emph{Szondi Personality Profile Space,} that is,
	the (inclusion-ordered, Boolean) powerset algebra \cite{DaveyPriestley} on \SPPts\ 
		(the set of all subsets of \SPPts).
\end{definition}
\noindent
As an example of an SPP,
	consider the \emph{norm profile} for the Szondi-test \cite{Szondi:ETD:Band1}:
		$$((\F{h}{,\p}), (\F{s}{,\p}), (\F{e}{,\m}), (\F{hy}{,\m}), 
				 (\F{k}{,\m}), (\F{p}{,\m}), (\F{d}{,\p}), (\F{m}{,\p}))$$
Spelled out, 
	this norm profile describes the personality of a human being who 
		approves of physical love,
		has a proactive attitude,
		has unethical but moral behaviour,
		wants to have and be less, and 
		is unfaithful and dependent.

We conclude this subsection with the definition of our interlingua LPL.
\begin{definition}[The Logical Pivot Language]
	Let 
		$$\atoms=\{\,\F{h}{s_{1}}, \F{s}{s_{2}}, \F{e}{s_{3}}, \F{hy}{s_{4}}, \F{k}{s_{5}}, \F{p}{s_{6}}, \F{d}{s_{7}}, \F{m}{s_{8}}  \mid  
					s_{1},\ldots,s_{8}\in\signatures\,\}$$
	be our set of atomic logical formulas, and 
		$\LPL(\atoms)$ the classical propositional language over $\atoms$, that is, 
		the set of sentences constructed from the elements in $\atoms$ and
			the classical propositional connectives 
				$\neg$ (negation, pronounced ``not''),
				$\land$ (conjunction, pronounced ``and''),
				$\lor$ (disjunction, pronounced ``or''), etc.
	
	Then,
		$$\LPLbis=\langle\,\LPL(\atoms),\Rightarrow\,\rangle$$
	defines our \emph{logical pivot language,} with
		$\Rightarrow$ being logical consequence.
		
	Logical equivalence $\equiv$ is defined in terms of $\Rightarrow$ such that 
		for every $\phi,\varphi\in\LPL(\atoms)$, 
			$\phi\equiv\varphi$ by definition if and only if 
				$\phi\Rightarrow\varphi$ and $\varphi\Rightarrow\phi$.
\end{definition}

\subsection{Mappings between structures}\label{section:MappingsAndMorphisms}
In this section, 
	we present 
		the defining mathematical mappings for 
			the concrete translation $\rightG{}$ of $\MBTIps$ to $\SPPps$ via $\LPLbis$ and 
			the abstract translation $\leftG{}$ of $\SPPps$ back to $\MBTIps$ again via $\LPLbis$ by 
				means of the auxiliary mappings $\rightI$ and $\leftI$.
We also prove that the ordered pair $(\,\rightG{},\leftG{}\,)$ is a Galois-connection, as promised.
\begin{definition}[Mappings]\label{definition:MappingsAndMorphisms}
	Let the mapping (total function)
		\begin{itemize}
			\item $\rightI$ be defined 
					in the function space $(\MBTIts\to\LPL(\atoms))$ as in Table~\ref{table:MBTItoLPL} and 
\begin{table}
\centering
\caption{Translating $\MBTI$ and $\MBTIts$ to $\LPL(\atoms)$}
\medskip
\resizebox{!}{0.48\textheight}{$\begin{array}{|@{\quad}r@{\ \ }c@{\ \ }l@{\quad}|}
	\hline
	\multicolumn{3}{|c|}{}\\[-3\jot]
	\rightI(\MBE) &=& \F{hy}{\p}\lor\F{hy}{\pb}\lor\F{hy}{\pbb}\lor\F{hy}{\pbbb}\lor\F{hy}{\pmub}\\[\jot]
	\rightI(\MBI) &=& \F{hy}{\m}\lor\F{hy}{\mb}\lor\F{hy}{\mbb}\lor\F{hy}{\mbbb}\lor\F{hy}{\pmlb}\\[\jot]
	\hline
	&&\\[-3\jot]
	\rightI(\MBF) &=& (\F{h}{\p}\lor\F{h}{\pm}\lor\F{h}{\pmlb})\land(\F{p}{\m}\lor\F{p}{\pm}\lor\F{p}{\pmub})\\[\jot]
	\rightI_{!}(\MBF) &=& \begin{array}[t]{@{}l@{}}
						(\F{h}{\pb}\lor\F{h}{\pbb}\lor\F{h}{\pbbb}\lor\F{h}{\pmub})\,\land\\
						(\F{p}{\mb}\lor\F{p}{\mbb}\lor\F{p}{\mbbb}\lor\F{p}{\pmlb})
					\end{array}\\[5\jot]
	\rightI(\MBT) &=& \F{k}{\m}\lor\F{k}{\pm}\lor\F{k}{\pmub}\\[\jot]
	\rightI_{!}(\MBT) &=& \F{k}{\mb}\lor\F{k}{\mbb}\lor\F{k}{\mbbb}\lor\F{k}{\pmlb}\\[\jot]
	\hline
	&&\\[-3\jot]
	\rightI(\MBN) &=& (\F{k}{\p}\lor\F{k}{\pm}\lor\F{k}{\pmlb})\land
					(\F{p}{\p}\lor\F{p}{\pm}\lor\F{p}{\pmlb})\\[\jot]
	\rightI_{!}(\MBN) &=& \begin{array}[t]{@{}l@{}}
						(\F{k}{\pb}\lor\F{k}{\pbb}\lor\F{k}{\pbbb}\lor\F{k}{\pmub})\,\land\\
						(\F{p}{\pb}\lor\F{p}{\pbb}\lor\F{p}{\pbbb}\lor\F{p}{\pmub})
					\end{array}\\[5\jot]
	\rightI(\MBS) &=& \begin{array}[t]{@{}l@{}l@{}}
					(&\F{k}{\p}\lor\F{k}{\pm}\lor\F{k}{\pmlb})\,\land\\
					(&(\F{h}{\p}\lor\F{e}{\m}\lor\F{hy}{\m}\lor\F{d}{\p}\lor\F{m}{\p})\,\lor\\
					 &(\F{h}{\pm}\lor\F{e}{\pm}\lor\F{hy}{\pm}\lor\F{d}{\pm}\lor\F{m}{\pm})\,\lor\\
					 &(\F{h}{\pmlb}\lor\F{e}{\pmub}\lor\F{hy}{\pmub}\lor\F{d}{\pmlb}\lor\F{m}{\pmlb}))
					\end{array}\\[13\jot]
	\rightI_{!}(\MBS) &=& \begin{array}[t]{@{}l@{}l@{}}
						(&\F{k}{\pb}\lor\F{k}{\pbb}\lor\F{k}{\pbbb}\lor\F{k}{\pmub})\,\land\\
						(&(\F{h}{\pb}\lor\F{e}{\mb}\lor\F{hy}{\mb}\lor\F{d}{\pb}\lor\F{m}{\pb})\,\lor\\
						 &(\F{h}{\pbb}\lor\F{e}{\mbb}\lor\F{hy}{\mbb}\lor\F{d}{\pbb}\lor\F{m}{\pbb})\,\lor\\
						 &(\F{h}{\pbbb}\lor\F{e}{\mbbb}\lor\F{hy}{\mbbb}\lor\F{d}{\pbbb}\lor\F{m}{\pbbb})\,\lor\\
						 &(\F{h}{\pmub}\lor\F{e}{\pmlb}\lor\F{hy}{\pmlb}\lor\F{d}{\pmub}\lor\F{m}{\pmub}))
					\end{array}\\[17\jot]
	\hline
	\hline	
	&&\\[-3\jot]
	\rightI(\ISTJ) &=& \rightI(\MBI)\land \rightI_{!}(\MBS)\land \rightI(\MBT)\\[\jot]
	 \rightI(\ISFJ) &=&  \rightI(\MBI)\land \rightI_{!}(\MBS)\land \rightI(\MBF)\\[\jot]
	 \rightI(\INFJ) &=&  \rightI(\MBI)\land \rightI_{!}(\MBN)\land \rightI(\MBF)\\[\jot]
	 \rightI(\INTJ) &=&  \rightI(\MBI)\land \rightI_{!}(\MBN)\land \rightI(\MBT)\\[\jot]
	 \rightI(\ISTP) &=&  \rightI(\MBI)\land \rightI(\MBS)\land \rightI_{!}(\MBT)\\[\jot]
	 \rightI(\ISFP) &=&  \rightI(\MBI)\land \rightI(\MBS)\land \rightI_{!}(\MBF)\\[\jot]
	 \rightI(\INFP) &=&  \rightI(\MBI)\land \rightI(\MBN)\land \rightI_{!}(\MBF)\\[\jot]
	 \rightI(\INTP) &=&  \rightI(\MBI)\land \rightI(\MBN)\land \rightI_{!}(\MBT)\\[\jot]
	 \rightI(\ESTP) &=&  \rightI(\MBE)\land \rightI_{!}(\MBS)\land \rightI(\MBT)\\[\jot]
	 \rightI(\ESFP) &=&  \rightI(\MBE)\land \rightI_{!}(\MBS)\land \rightI(\MBF)\\[\jot]
	 \rightI(\ENFP) &=&  \rightI(\MBE)\land \rightI_{!}(\MBN)\land \rightI(\MBF)\\[\jot]
	 \rightI(\ENTP) &=&  \rightI(\MBE)\land \rightI_{!}(\MBN)\land \rightI(\MBT)\\[\jot]
	 \rightI(\ESTJ) &=&  \rightI(\MBE)\land \rightI(\MBS)\land \rightI_{!}(\MBT)\\[\jot]
	 \rightI(\ESFJ) &=&  \rightI(\MBE)\land \rightI(\MBS)\land \rightI_{!}(\MBF)\\[\jot]
	 \rightI(\ENFJ) &=&  \rightI(\MBE)\land \rightI(\MBN)\land \rightI_{!}(\MBF)\\[\jot]
	 \rightI(\ENTJ) &=&  \rightI(\MBE)\land \rightI(\MBN)\land \rightI_{!}(\MBT)\\[\jot]
	\hline
\end{array}$}
\label{table:MBTItoLPL}
\end{table}
					in the function space $(2^{\MBTIts}\to\LPL(\atoms))$ such that for every $I\in2^{\MBTIts}$, 
						$$\rightI(I) = \bigwedge\{\,\rightI(i) \mid i\in I\,\}\,;$$
			\item $\leftI$ be defined in the function space $(\SPPts\to\LPL(\atoms))$ such that 
					$$\begin{array}{@{}r@{}}
						 \leftI(((\F{h}{,s_{1}}), (\F{s}{,s_{2}}), (\F{e}{,s_{3}}), (\F{hy}{,s_{4}}), 
			(\F{k}{,s_{5}}), (\F{p}{,s_{6}}), (\F{d}{,s_{7}}), (\F{m}{,s_{8}})))=\\
						\F{h}{s_{1}}\land\F{s}{s_{2}}\land\F{e}{s_{3}}\land\F{hy}{s_{4}}\land\F{k}{s_{5}}\land\F{p}{s_{6}}\land\F{d}{s_{7}}\land\F{m}{s_{8}}
					\end{array}$$
					and in the function space $(2^{\SPPts}\to\LPL(\atoms))$ such that for every $P\in2^{\SPPts}$,
						$$ \leftI(P) = \bigvee\{\, \leftI(p) \mid p\in P\,\}\,.$$
		\end{itemize}
	Then, the mapping  
		\begin{itemize}
			\item $\rightG{}:\MBTIps\to\SPPps$ defined such that for every $I\in2^{\MBTIts}$, 
					$$\rightG{I} = \{\,p\in\SPPts \mid \leftI(p)\Rightarrow\rightI(I)\,\}$$
					is the so-called \emph{right polarity} and
			\item $\leftG{}:\SPPps\to\MBTIps$ defined such that for every $P\in2^{\SPPts}$, 
					$$\leftG{P} = \{\,i\in\MBTIts \mid \leftI(P)\Rightarrow\rightI(i)\,\}$$
					is the so-called \emph{left polarity} of the ordered pair $(\,\rightG{},\leftG{}\,)$.
		\end{itemize}		
\end{definition}
\noindent 
Spelled out, 
	(1) the result of 
			applying the mapping $\rightI$ to 
				a set $I$ of MBTIs $i$ as defined in Definition~\ref{definition:MappingsAndMorphisms} is 
					the conjunction of the results of 
						applying $\rightI$ to 
							each one of these $i$ as defined in Table~\ref{table:MBTItoLPL};
	(2) the result of 
			applying the mapping $\leftI$ to 
				a set $P$ of SPPs $p$ as defined in Definition~\ref{definition:MappingsAndMorphisms} is 
					the disjunction of the results of 
						applying $\leftI$ to 
							each one of these $p$, which 
								simply is the conjunction of 
									all signed factors in $p$ taken each one as an atomic proposition;
	(3)	the result of 
			applying the mapping $\rightG{}$ to
				a set $I$ of MBTIs is 
					the set of all those SPPs $p$ whose 
						mapping under $\leftI$ implies the mapping of $I$ under $\rightI$;
	(4) the result of 
			applying the mapping $\leftG{}$ to
				a set $P$ of SPPs is 
					the set of all those MBTIs $i$ whose 
						mapping under $\rightI$ is implied by the mapping of $P$ under $\leftI$.	
Thus from a computer science perspective \cite[Section~7.35]{DaveyPriestley}, 
	MBTIs are specifications of SPPs and 
	SPPs are implementations or refinements of MBTIs.
The Galois-connection then connects correct implementations to their respective specification by 
	stipulating that a correct implementation imply its specification.
By convention,
	$\bigwedge\emptyset=\top$ and $\bigvee\emptyset=\bot$\,, that is,
		the conjunction over the empty set $\emptyset$ is tautological truth $\top$\,, and 
		the disjunction over $\emptyset$ is tautological falsehood $\bot$\,, respectively.

Note that an example of an SPP that 
	maps to the empty set under $\leftG{}$ happens to be the Szondi norm profile mentioned before, because 
		its mapping under $\leftI$ 
	$$\begin{array}{r}
		\leftI(((\F{h}{,\p}), (\F{s}{,\p}), (\F{e}{,\m}), (\F{hy}{,\m}), 
				 (\F{k}{,\m}), (\F{p}{,\m}), (\F{d}{,\p}), (\F{m}{,\p})))=\\
				 	\F{h}{\p}\land\F{s}{\p}\land\F{e}{\m}\land\F{hy}{\m}\land\F{k}{\m}\land\F{p}{\m}\land\F{d}{\p}\land\F{m}{\p}\,,
	\end{array}$$
does not contain any (individually) dominant factor
	(all factors are simply either positive or negative, and thus are without quanta).
So it does not imply the mapping of any MBTI under $\rightI$, which 
	requires a dominant factor, as we have 
		alluded to in the introduction with the dominance flag, and 
		are going to explain now by 
			spelling out the translation of MBTIs to SPPs, see Table~\ref{table:MBTItoLPL}.
There,
	dominance is indicated by 
		a quantum subscript of the translation $\rightI$.
As can be seen,
	\MBJ\ and \MBP\ merely act as flag values, which 
		indicate which one of the judgment or perception faculties is 
			the dominant faculty \emph{for dealing with the outer world,} 
				(that is, the observably dominant faculty) and 
					this as a function of extro-/introversion.
The rule is that
	extroverts do show their dominant faculty for dealing with the outer world, whereas 
	introverts do not \cite[Page~14]{GiftsDiffering}. 
So 
	with judging (\MBJ) introverts (\MBI), a perceptive faculty shows up as dominant,
	with perceiving (\MBP) introverts (\MBI), a judging faculty, 
	with perceiving (\MBP) extroverts (\MBE), a perceptive faculty, and 
	with judging (\MBJ) extroverts (\MBE), a judging faculty.
As can be seen in Table~\ref{table:MBTItoLPL},
	our interpretation of extroversion is a positive tendency of Szondi's factor \F{hy}{} 
		(immoral behaviour: being seen, showing off).
Whereas our interpretation of introversion is a negative tendency thereof 
		(moral behaviour: seeing, hiding).
Our readers might want to stipulate additional constraints for their own interpretation, but
	must take care not to create conflicts.
Consistency is a necessary condition for the faithfulness of the translation $\rightI$\,!
(It is also one for the faithfulness of the translation $\leftI$, but there it is obvious.)
\begin{fact}[Consistency of the translation $\rightI$]\ 
	\begin{enumerate}
		\item For every $b\in\{\MBE,\MBI\}$ and $b'\in\MBTI\setminus\{\MBE,\MBI\}$,
				$\rightI(b)\land\rightI(b')\not\equiv\bot$\,.
		\item For every $b\in\{\MBF,\MBT\}$,
			\begin{itemize}
				\item $\rightI(b)\land\rightI_{!}(N)\not\equiv\bot$ and 
					$\rightI(b)\land\rightI_{!}(S)\not\equiv\bot$\,;
				\item $\rightI_{!}(b)\land\rightI(N)\not\equiv\bot$ and 
					$\rightI_{!}(b)\land\rightI(S)\not\equiv\bot$\,.
			\end{itemize}
	\end{enumerate}
\end{fact}
\begin{proof}
	By inspection of Table~\ref{table:MBTItoLPL}.
\end{proof}
\noindent
Our interpretations of the remaining faculties follow a simple \emph{generating pattern:}
	\begin{itemize}
		\item for a non-dominant positive factor $f$, 
				the pattern is $\F{\text{$f$}}{\p}\lor\F{\text{$f$}}{\pm}\lor\F{\text{$f$}}{\pmlb}$\,;
		\item for a non-dominant negative factor $f$, 
				it is $\F{\text{$f$}}{\m}\lor\F{\text{$f$}}{\pm}\lor\F{\text{$f$}}{\pmub}$\,;
		\item for a dominant positive factor $f$, 
				it is $\F{\text{$f$}}{\pb}\lor\F{\text{$f$}}{\pbb}\lor\F{\text{$f$}}{\pbbb}\lor\F{\text{$f$}}{\pmub}$\,;
		\item for a dominant negative factor $f$, 
				it is $\F{\text{$f$}}{\mb}\lor\F{\text{$f$}}{\mbb}\lor\F{\text{$f$}}{\mbbb}\lor\F{\text{$f$}}{\pmlb}$\,.
	\end{itemize}
As can be seen in Table~\ref{table:MBTItoLPL},
	our interpretation of feeling is the conjunction of 
		personal warmth (\F{h}{\p}) and 
		empathy (\F{p}{\m}), which
			\cite{MyersBriggs} stipulate as the characteristic properties of this faculty.
Our interpretation of thinking is simply its corresponding factor (\F{k}{\m}) in Szondi's system.
Our interpretation of the two perceptive faculties contains as a conjunct the factor (\F{k}{\p}), which 
	corresponds to perception in Szondi's system.
Our interpretation of intuition then further contains its 
	corresponding factor (\F{p}{\p}) in Szondi's system.
Finally,
	our interpretation of sensing further contains the
		disjunction of all those factors that correspond to the human senses in Szondi's system, namely:
			touching (\F{h}{\p}), 
			hearing (\F{e}{\m}), 
			seeing (\F{hy}{\m}), 
			smelling (\F{d}{\p}), and
			tasting (\F{m}{\p}).

Our readers might be interested in comparing our interpreting mapping of MBTIs with 
	D.W.~Keirsey's \cite{PleaseUnderstandMe}: for him, 
		\ISTJ\ maps to the Inspector,
		\ISFJ\ to the Protector,
		\INFJ\ to the Counselor,
		\INTJ\ to the Mastermind,
		\ISTP\ to the Crafter,
		\ISFP\ to the Composer,
		\INFP\ to the Healer,
		\INTP\ to the Architect,
		\ESTP\ to the Promoter,
		\ESFP\ to the Performer,
		\ENFP\ to the Champion,
		\ENTP\ to the Inventor,
		\ESTJ\ to the Supervisor,
		\ESFJ\ to the Provider,
		\ENFJ\ to the Teacher, and
		\ENTJ\ to the Fieldmarshal.

We now prove in two intermediate steps that 
	the ordered pair $(\,\rightG{},\leftG{}\,)$ is indeed a Galois-connection.
The first step is the following announced fact, from which 
	the second step, Lemma~\ref{lemma:Properties}, follows, from which in turn 
		the desired result, Theorem~\ref{theorem:Galois}, then follows---easily.
As announced,
	all that our readers need to check on their own analog of our mapping $\rightI$ is 
		that it has the property stated as Fact~\ref{fact:FactsAboutip}.1.
Their own Galois-connection is then automatically induced.
\begin{fact}[Some facts about $\rightI$ and $\leftI$]\label{fact:FactsAboutip}\ 
	\begin{enumerate}
		\item if $I\subseteq I'$ then $\rightI(I')\Rightarrow\rightI(I)$
		\item if $P\subseteq P'$ then $\leftI(P)\Rightarrow\leftI(P')$
		\item The functions $\rightI$ and $\leftI$ are \emph{injective} but not surjective.
	\end{enumerate}
\end{fact}
\begin{proof}
	By inspection of Definition~\ref{definition:MappingsAndMorphisms} and Table~\ref{table:MBTItoLPL}.
\end{proof}
\noindent
We need 
	Fact~\ref{fact:FactsAboutip}.1 and \ref{fact:FactsAboutip}.2 but 
	not Fact~\ref{fact:FactsAboutip}.3 in the following development.
Therefor, note the two macro-definitions 
	$\rightleftG{}:=\rightG{}\circ\leftG{}$ and 
	$\leftrightG{}:=\leftG{}\circ\rightG{}$ with 
		$\circ$ being function composition, as usual (from right to left, as usual too).
\begin{lemma}[Some useful properties of $\rightG{}$ and $\leftG{}$]\label{lemma:Properties}\ 
	\begin{enumerate}
		\item if $I\subseteq I'$ then $\rightG{I'}\subseteq\rightG{I}$\quad(\;$\rightG{}$ is antitone)
		\item if $P\subseteq P'$ then $\leftG{P'}\subseteq\leftG{P}$\quad(\,$\leftG{}$ is antitone)
		\item $P\subseteq\rightG{(\leftG{P})}$\quad(\;$\rightleftG{}$ is inflationary)
		\item $I\subseteq\leftG{(\rightG{I})}$\quad(\,$\leftrightG{}$ is inflationary)
	\end{enumerate}
\end{lemma}
\begin{proof}
	For (1), 
		let $I,I'\in2^{\MBTIts}$ and 
		suppose that $I\subseteq I'$.
	Hence 
		$\rightI(I')\Rightarrow\rightI(I)$ by Fact~\ref{fact:FactsAboutip}.1.
	Further suppose that 
		$p\in\rightG{I'}$.
	By definition,
		$\rightG{I'} = \{\,p\in\SPPts \mid \leftI(p)\Rightarrow\rightI(I')\,\}$.
	Hence $\leftI(p)\Rightarrow\rightI(I')$.
	Hence $\leftI(p)\Rightarrow\rightI(I)$ by transitivity.
	By definition,
		$\rightG{I} = \{\,p\in\SPPts \mid \leftI(p)\Rightarrow\rightI(I)\,\}$.	
	Hence $p\in\rightG{I}$.
	Thus $\rightG{I'}\subseteq\rightG{I}$.
	
	For (2), 
		let $P,P'\in2^{\SPPts}$ and 
		suppose that $P\subseteq P'$.
	Hence 
		$\leftI(P)\Rightarrow\leftI(P')$ by Fact~\ref{fact:FactsAboutip}.2.
	Further suppose that 
		$i\in\leftG{P'}$.
	By definition,
		$\leftG{P'} = \{\,i\in\MBTIts \mid \leftI(P')\Rightarrow\rightI(i)\,\}$.
	Hence $\leftI(P')\Rightarrow\rightI(i)$.
	Hence $\leftI(P)\Rightarrow\rightI(i)$ by transitivity.
	By definition,
		$\leftG{P} = \{\,i\in\MBTIts \mid \leftI(P)\Rightarrow\rightI(i)\,\}$.	
	Hence $i\in\leftG{P}$.
	Thus $\leftG{P'}\subseteq\leftG{P}$.	
	
	For (3), consider:
		\begin{enumerate}
			\item \quad$p\in P$\hfill hypothesis\\[-5\jot]
			\item \quad$\{p\}\subseteq P$\hfill 1\\[-5\jot]
			\item \quad$\leftI(p)\Rightarrow\leftI(P)$\hfill 2, Fact~\ref{fact:FactsAboutip}.2\\[-5\jot]
			\item \qquad $\leftI(p)$ is true\hfill hypothesis\\[-5\jot]
			\item \qquad $\leftI(P)$ is true\hfill 3, 4\\[-5\jot]
			\item \qquad\quad $\phi\in\{\,\rightI(i)\mid\leftI(P)\Rightarrow\rightI(i)\,\}$\hfill hypothesis\\[-5\jot]
			\item \qquad\quad there is $i$ s.t.\ $\phi=\rightI(i)$ and $\leftI(P)\Rightarrow\rightI(i)$\hfill 6\\[-5\jot]
			\item \qquad\qquad $\phi=\rightI(i)$ and $\leftI(P)\Rightarrow\rightI(i)$\hfill hypothesis\\[-5\jot]
			\item \qquad\qquad $\leftI(P)\Rightarrow\rightI(i)$\hfill 8\\[-5\jot]
			\item \qquad\qquad $\rightI(i)$ is true\hfill 5, 9\\[-5\jot]
			\item \qquad\qquad $\phi=\rightI(i)$\hfill 8\\[-5\jot]
			\item \qquad\qquad $\phi$ is true\hfill 10, 11\\[-5\jot]
			\item \qquad\quad $\phi$ is true\hfill 7, 8--12\\[-5\jot]
			\item \qquad for every $\phi\in\{\,\rightI(i)\mid\leftI(P)\Rightarrow\rightI(i)\,\}$, $\phi$ is true\hfill 6--13\\[-5\jot]
			\item \qquad $\bigwedge\{\,\rightI(i)\mid\leftI(P)\Rightarrow\rightI(i)\,\}$ is true\hfill 14\\[-5\jot]
			\item \qquad $\bigwedge\{\,\rightI(i)\mid i\in\{\,i\in\MBTIts\mid\leftI(P)\Rightarrow\rightI(i)\,\}\}$ is true\hfill 15\\[-5\jot]
			\item \qquad $\bigwedge\{\,\rightI(i)\mid i\in\leftG{P}\,\}$ is true\hfill 16\\[-5\jot]
			\item \qquad $\rightI(\leftG{P})$ is true\hfill 17\\[-5\jot]
			\item \quad $\leftI(p)\Rightarrow\rightI(\leftG{P})$\hfill 4--18\\[-5\jot]
			\item \quad $p\in\{\,p\in\SPPts\mid\leftI(p)\Rightarrow\rightI(\leftG{P})\,\}$\hfill 19\\[-5\jot]
			\item \quad $p\in\rightG{(\leftG{P})}$\hfill 20\\[-5\jot]
			\item $P\subseteq\rightG{(\leftG{P})}$\hfill 1--21.
		\end{enumerate}
	For (4), consider:
		\begin{enumerate}
			\item \quad$i\in I$\hfill hypothesis\\[-5\jot]
			\item \quad$\{i\}\subseteq I$\hfill 1\\[-5\jot]
			\item \quad$\rightI(I)\Rightarrow\rightI(i)$\hfill 2, Fact~\ref{fact:FactsAboutip}.1\\[-5\jot]
			\item \qquad $\leftI(\rightG{I})$ is true\hfill hypothesis\\[-5\jot]
			\item \qquad $\bigvee\{\,\leftI(p)\mid p\in\rightG{I}\,\}$ is true\hfill 4\\[-5\jot]
			\item \qquad $\bigvee\{\,\leftI(p)\mid p\in\{\,p\in\SPPts\mid\leftI(p)\Rightarrow\rightI(I)\,\}\,\}$ is true\hfill 5\\[-5\jot]
			\item \qquad $\bigvee\{\,\leftI(p)\mid \leftI(p)\Rightarrow\rightI(I)\,\}$ is true\hfill 6\\[-5\jot]			
			\item \qquad there is $p$ s.t.\ $\leftI(p)\Rightarrow\rightI(I)$ and $\leftI(p)$ is true\hfill 7\\[-5\jot]
			\item \qquad\quad $\leftI(p)\Rightarrow\rightI(I)$ and $\leftI(p)$ is true\hfill hypothesis\\[-5\jot]
			\item \qquad\quad $\rightI(I)$ is true\hfill 9\\[-5\jot]
			\item \qquad\quad $\rightI(i)$ is true\hfill 3, 10\\[-5\jot]
			\item \qquad $\rightI(i)$ is true\hfill 8, 9--11\\[-5\jot]
			\item \quad $\leftI(\rightG{I})\Rightarrow\rightI(i)$\hfill 4--12\\[-5\jot]
			\item \quad $i\in\{\,i\in\MBTIts\mid\leftI(\rightG{I})\Rightarrow\rightI(i)\,\}$\hfill 13\\[-5\jot]
			\item \quad $i\in\leftG{(\rightG{I})}$\hfill 14\\[-5\jot]
			\item $I\subseteq\leftG{(\rightG{I})}$\hfill 1--15.
		\end{enumerate}

\end{proof}
\noindent
We are ready for making the final step.
\begin{theorem}[The Galois-connection property of $(\,\rightG{},\leftG{}\,)$]\label{theorem:Galois}
	The ordered pair $(\,\rightG{},\leftG{}\,)$ is an \emph{antitone} or \emph{order-reversing Galois-connection} between $\MBTIps$ and $\SPPps$.
	That is, 
		for every $I\in2^{\MBTIts}$ and $P\in2^{\SPPts}$,
			$$\text{$P\subseteq\rightG{I}$ if and only if $I\subseteq\leftG{P}$.}$$
\end{theorem}
\begin{proof}
	Let $I\in2^{\MBTIts}$ and $P\in2^{\SPPts}$.
	Suppose that $P\subseteq\rightG{I}$.
	Hence $\leftG{(\rightG{I})}\subseteq\leftG{P}$ by Lemma~\ref{lemma:Properties}.2.
	Further,
		$I\subseteq\leftG{(\rightG{I})}$ by Lemma~\ref{lemma:Properties}.4.
	Hence $I\subseteq\leftG{P}$ by transitivity.
	Conversely suppose that $I\subseteq\leftG{P}$.
	Hence $\rightG{(\leftG{P})}\subseteq\rightG{I}$ by Lemma~\ref{lemma:Properties}.1.
	Further,
		$P\subseteq\rightG{(\leftG{P})}$ by Lemma~\ref{lemma:Properties}.3.
	Hence $P\subseteq\rightG{I}$.
\end{proof}
\noindent
Thus from a computer science perspective \cite[Section~7.35]{DaveyPriestley}, 
	smaller (larger) sets of MBTIs and thus less (more) restrictive specifications correspond to 
	larger (smaller) sets of SPPs and thus more (fewer) possible implementations.

Note that Galois-connections are 
	connected to \emph{residuated mappings} \cite{LatticesAndOrderedAlgebraicStructures}.
Further, 
	natural notions of equivalence on $\MBTIps$ and $\SPPps$ are given by 
		the \emph{kernels} of $\rightG{}$ and $\leftG{}$, respectively, which are, by definition:
$$\begin{array}{rcl}
	I\equiv I' &\text{if and only if}& \rightG{I}=\rightG{I'}\;;\\[\jot]
	P\equiv P' &\text{if and only if}& \leftG{P}=\leftG{P'}\,.
\end{array}$$

\begin{proposition}[The efficient computability of $(\,\rightG{},\leftG{}\,)$]\ 
	\begin{enumerate}
		\item Given $I\in2^{\MBTIts}$, $\rightG{I}$ is efficiently computable.
		\item Given $P\in2^{\SPPts}$, $\leftG{P}$ is efficiently computable.
	\end{enumerate}
\end{proposition}
\begin{proof}
	Even a relatively brute-force approach is feasible with 
		today's laptop computers, which
			have a (high-speed) random access memory of several giga ($10^{9}$) bytes and
			a processor power of several giga instructions per second.
	Moreover in psychological practice, 
		the sets $I$ and $P$ will usually be singletons of 
			either only one type indicator 
			or only one personality profile, respectively.
	\begin{itemize}		
		\item Given $I\in2^{\MBTIts}$ (thus $0\leq|I|\leq16$), 
				we precompute the list of all SPPs, which
					contains $16^{8}$ entries, once and for all $I$, and 
						store the list 
							on a computer hard disk and then 
							load it into the (fast) random access memory of the computer.
			Then we model-check the formula $\rightI(I)$, which 
				can also be precomputed, against each entry $p$ in the list.
			That is,
				we check whether a given model $p$ satisfies (makes true) the given $\rightI(I)$, and	
				collect up into the result set all those $p$ that do satisfy $\rightI(I)$.
			It is well-known that model-checking a propositional formula, such as $\rightI(I)$, 
				takes only a polynomial number of computation steps in the size of the formula, and 
					thus only a polynomial number in the size of $I$ (being at most 16).
					
			Of course, 
				this computation can be done once and for all of the $2^{16}$ possible sets $I$, and
					the results stored on a hard disk for faster, later look-up.
		\item Given $P\in2^{\SPPts}$ (thus $0\leq|P|\leq16^{8}$),
			we model-check 
				each $p$ in $P$ against the (pre-computable) mapping $\rightI(i)$ of each $i$ of the 16 MBTIs, and 
				collect up into the result set all those $i$ whose mapping $\rightI(i)$ satisfies $p$.
	\end{itemize}
\end{proof}
\noindent
Of course, 
	optimisations of the computation procedure given in the previous proof are possible, but 
		we consider them as not sufficiently interesting implementation details.

\section{Conclusion}
We have proposed a computable Galois-connection between 
	Myers-Briggs Type Indicators and 
	Szondi's personality profiles as
		promised in the abstract.
In addition, 
	we have proposed a simple methodology for 
		generating other such Galois-connections, 
			including Galois-connections 
				not only between this pair of spaces of personality-test result values 
				but also between other such pairs.

\paragraph{Acknowledgements} 
I thank Danilo Diedrichs for proof-reading this article.
The \LaTeX-package TikZ was helpful for graph drawing.

\bibliographystyle{plain}

\begin{thebibliography}{10}

\bibitem{LatticesAndOrderedAlgebraicStructures}
T.S. Blyth.
\newblock {\em Lattices and Ordered Algebraic Structures}.
\newblock Springer, 2005.

\bibitem{DaveyPriestley}
B.A. Davey and H.A. Priestley.
\newblock {\em Introduction to Lattices and Order}.
\newblock Cambridge University Press, 2nd edition, 1990 (2002).

\bibitem{PsychologicalTypes}
C.G. Jung.
\newblock {\em Psychological Types}, volume~6 of {\em The Collected Works of
  C.G. Jung}.
\newblock Princeton University Press, 1971.

\bibitem{PleaseUnderstandMe}
D.~Keirsey.
\newblock {\em Please Understand Me II}.
\newblock Prometheus Nemesis Book Company, 1984 (1998).

\bibitem{Kenmo:Szondi}
R.~Kenmo.
\newblock {\em Let the Personality Bloom}.
\newblock Humankonsult, 2009.

\bibitem{SzondiTestWebApp}
S.~Kramer.
\newblock \url{www.szondi-test.ch}, 2014.
\newblock forthcoming.

\bibitem{MyersBriggs}
I.~Briggs Myers and M.H. McCaulley.
\newblock {\em Manual: A Guide to the Development and Use of the Myers-Briggs
  Type Indicator}.
\newblock Consulting Psychologists Press, 2nd edition, 1985.

\bibitem{GiftsDiffering}
I.~Briggs Myers and P.B. Myers.
\newblock {\em Gifts Differing: Understanding Personality Type}.
\newblock Consulting Psychologists Press, 1980 (1995).

\bibitem{Szondi:ETD:Band1}
L.~Szondi.
\newblock {\em Lehrbuch der Experimentellen Triebdiagnostik}, volume I:
  Text-Band.
\newblock Hans Huber, 3rd edition, 1972.

\bibitem{Szondi:IchAnalyse}
L.~Szondi.
\newblock {\em Ich-Analyse: Die Grundlage zur Vereinigung der
  Tiefenpsychologie}.
\newblock Hans Huber, 1999.
\newblock English translation:
  \url{https://sites.google.com/site/ajohnstontranslationsofszondi/}.

\bibitem{Szondi:Schicksalsanalyse}
L.~Szondi.
\newblock {\em Schicksalsanalyse: Wahl in Liebe, Freundschaft, Beruf, Krankheit
  und Tod}.
\newblock Schwabe, 4th edition, 2004.

\end{thebibliography}

\end{document}